\documentclass[useAMS,usenatbib]{mn2e}
\usepackage{times}
\usepackage{graphicx}

\def\aap{A\&A}%
\def\aaps{A\&AS}%
\def\actaa{Acta Astronomica}%
\def\aj{AJ}%
\def\apj{ApJ}%
\def\apjs{ApJ Suppl.~Series}%
\def\apss{Astroph.~\& Space Sci.}
\def\mnras{MNRAS}%
\def\nat{Nature}%
\def\pasp{PASP}%
\def\ssr{Space Science Rev.}%

\title[Life after eruption. I.]{Life after eruption -- I. 
Spectroscopic observations of ten nova candidates}
\author[C. Tappert et al.]%
{
C. Tappert,$^{1}$\thanks{E-mail: claus.tappert@uv.cl} 
A. Ederoclite,$^{2}$
R.~E. Mennickent,$^{3}$
L. Schmidtobreick$^{4}$ and
N. Vogt$^{1}$%
\footnotemark[1]\thanks{Based on observations with ESO telescopes,
proposal numbers 083.D-0158, 086.D-0428, and 087.D-0323}\\
$^{1}$Dpto. de F\'{\i}sica y Astronom\'{\i}a, Universidad de Valpara\'{\i}so,
Avda. Gran Breta\~na 1111, Valpara\'{\i}so, Chile\\
$^{2}$Centro de Estudios de F\'{\i}sica del Cosmos de Arag\'on, Plaza San 
Juan 1, Planta 2, Teruel, E44001, Spain\\
$^{3}$Dpto. de Astronom\'{\i}a, Universidad de Concepci\'on, Casilla 160-C,
Concepci\'on, Chile\\
$^{4}$European Southern Observatory, Alonso de Cordova 3106, Santiago, Chile
}
\begin{document}

\date{Accepted. Received}

\pagerange{\pageref{firstpage}--\pageref{lastpage}} \pubyear{2011}

\maketitle

\label{firstpage}

\begin{abstract}
We have started a project to investigate the connection of post-novae with
the population of cataclysmic variables. Our first steps in this concern
improving the sample of known post-novae and their properties. Here we present
the recovery and/or confirmation of the old novae MT Cen, V812 Cen, 
V655 CrA, IL Nor, V2109 Oph, V909 Sgr, V2572 Sgr, and V728 Sco. Principal 
photometric and spectroscopic properties of these systems are discussed. We 
find that V909 Sgr is a probable magnetic CV, and that V728 Sco is a 
high-inclination system.  We furthermore suggest that the two candidate novae 
V734 Sco and V1310 Sgr have been misclassified and instead are Mira variables.
\end{abstract}


\section{Introduction}

A nova eruption in a cataclysmic binary star (CV) occurs as a thermonuclear
explosion on the surface of the white dwarf primary star once it has
accumulated a critical mass from its late-type, usually main-sequence,
companion. In the process of the eruption, material
is ejected into the interstellar medium. The mass of the shell typically
amounts to $10^{-5}$ to $10^{-4}~\mathrm{M_\odot}$
\citep[e.g.,][]{yaronetal05-1}. There is recent evidence that the mass
of the white dwarf increases in the course of the secular evolution of CVs
\citep*{zorotovicetal11-1}. Consequently, in nova eruptions less than the
accumulated mass is ejected. With typical mass-transfer rates of a few
$10^{-8}$ to $10^{-9}~\mathrm{M_\odot~yr^{-1}}$ \citep{townsley+gaensicke09-1}
the typical recurrence time of a nova eruption $>10^{3}~\mathrm{yr}$. This
distinguishes the classical novae from the recurrent novae that have much
shorter recurrence times in the order of decades, and in most cases have
a giant donor.

In between nova eruptions the binary is supposed to appear as a ``normal"
CV, i.e.~its behaviour is dominated by its current mass-transfer rate
and the magnetic field strength of the white dwarf \citep{vogt89-1}.
Furthermore, the hibernation model predicts that most of the time between
eruptions the system passes as a detached binary
\citep{sharaetal86-1,prialnik+shara86-1}. The proposed scenario here is that
the binary separation increases due to the 
mass lost from the white dwarf during the eruption.
The primary is heated by the thermonuclear runaway, thus in return heats the
companion star, driving it far out of thermal equilibrium, and so sustaining
a high mass-transfer rate in spite of the increased separation. The 
mass-transfer rate gradually decreases and eventually the donor can relax to
thermal equilibrium, losing contact to its Roche lobe, and the system becomes
detached. While there is still no observational evidence for the latter
part of this scenario, i.e.~the state of actual ``hibernation"
\citep[e.g.,][]{nayloretal92-1}, it is already well established that old novae 
are part of the CV class. For example, the system DQ Her (Nova Her 1934) is 
known as the prototype intermediate polar, 
while RR Pic (Nova Pic 1925) shows the characteristics of an SW Sex system 
\citep*{schmidtobreicketal03-4}. GK Per (Nova Per 1901) seems to be an
intermediate polar, revealing, decades after its nova eruption, the typical
behaviour of a dwarf nova, with semi-periodic outbursts \citep{simon02-4}.
Furthermore, the discovery of a nova shell around the dwarf nova Z Cam 
\citep{sharaetal07-1} proves that (at least some) systems discovered as CVs 
are also old novae. 

In spite of these, more or less isolated and exotic, cases, there is still only 
a fragmentary knowledge of the generally valid long-term behaviour of 
classical novae before and after their eruption. A study based on 97 
relatively well observed galactic novae by \citet{vogt90-1} revealed a  
decrease in brightness with a mean slope $21 \pm 6~\mathrm{mmag}$ per year
during the first 130 years after the eruption. Similar results were obtained by
\citet{duerbeck92-3} who derived, from a sample of nine very well covered 
cases, a mean decline rate of $10 \pm 3~\mathrm{mmag}$ per year, half 
a century after outburst. This is all our knowledge, at present.

Most classical novae have orbital periods between 3 and 6 hours
\citep{diaz+bruch97-1}. In this period range we find several other classes 
of CVs, in particular dwarf novae, magnetic CVs, and nova-like variables like 
SW Sex and UX UMa stars, most of them characterised by high mass-transfer 
rates. Do all these CVs suffer nova eruptions? Is the magnetic field of the 
white dwarf of any importance for the eruption recurrence time? And do 
post-novae really go into hibernation, i.e.~do CVs become detached for a time 
as a consequence of the nova eruption? We do not know, although answering 
these questions would be rather important for our understanding of CVs. The 
only way to progress here would be a more exhaustive investigation of the 
detailed behaviour of old novae decades and centuries after their outbursts. 
As a first step towards this goal, it is necessary to identify the remnants of 
past novae, especially those which had erupted long time ago. 

Our knowledge, in this respect, is largely incomplete. There are about 200 
confirmed or suspected novae with eruptions before
1980, but for less than half of them a spectrum of the post-nova has been
obtained. Only for 39 of these, a value for the orbital period is 
listed. Moreover, eight of these can be regarded as uncertain since they were 
obtained photometrically without the light curve presenting definite orbital 
features like eclipses or ellipsoidal variations, two more (DY Pup and DI Lac) 
are based on unpublished data \citep{downesetal05-1}, and the 5.714 d period 
of V1017 Sgr is marked as ``preliminary" \citep{sekiguchi92-3}. This leaves
a mere 28 old novae with a well-established orbital period.

To improve the current situation on systematic research on old novae we have
started a program to identify nova candidates with $U\!BV\!R$ photometry
via their specific colours and to confirm them with low-resolution spectroscopy
\citep{schmidtobreicketal03-5,schmidtobreicketal05-2}. For sufficiently bright 
systems, we furthermore plan to determine their orbital period by measuring 
radial velocities obtained with time-series spectroscopy. Since our
main interest regards the underlying CV we intend to limit our research to
novae where the nova shell is already sufficiently faint to provide only
a negligible contribution at least to the optical spectrum. The time scale
for the fading of the nova shell will be different for individual systems.
However, a literature research on well-known novae shows that after about
30 years the nebular lines have disappeared from the spectra in almost
all systems\footnote{For a counter-example see the nova HR Del that still
presents a significant contribution of nebular lines 40 years after the
eruption \citep*{friedjungetal10-2}}. In order to facilitate the selection, 
we therefore limit our study to novae that erupted before 1980. 

We here present the results for ten candidate old novae.

\section{Observations and reduction}

\begin{table*}
\begin{minipage}{13.7cm}
\caption[]{Log of observations.}
\label{obslog_tab}
\begin{tabular}{lllllll}
\hline\noalign{\smallskip}
object & R.A. (2000.0) & DEC (2000.0) & date & filter/grism 
& $t_\mathrm{exp}$ [s] & mag$^1$ \\
\hline\noalign{\smallskip}
MT Cen    & 11:44:00.24 & $-$60:33:35.7 & 2009-05-20 & $U\!BV\!R$  
& 1800/900/300/180 & 19.8\,$V$\\
          &             &               & 2011-02-16 & grism 11, slit 1.0 
& 3600             & 19.5\,$R$\\
V812 Cen  & 13:13:54.32 & $-$57:40:44.4 & 2009-05-21 & $U\!BV\!R$ 
& 1800/900/300/180 & 21.3\,$V$\\
          &             &               & 2009-05-22 & grism 4, slit 1.0 
& 5400             & 20.0\,$R$\\
V655 CrA  & 18:24:44.73 & $-$36:59:41.8 & 2009-05-23 & grism 4, slit 1.5 
& 1080             & 17.7\,$R$\\
IL Nor    & 15:29:23.00 & $-$50:35:00.4 & 2009-05-21 & $U\!BV\!R$ 
& 1800/900/300/180 & 19.0\,$V$\\
          &             &               & 2011-02-26 & grism 4, slit 1.0 
& 1800             & 18.9\,$R$\\
V2109 Oph & 17:24:16.04 & $-$24:36:50.2 & 2009-05-24 & grism 4, slit 1.0 
& 2700             & 19.7\,$R$\\
V909 Sgr  & 18:25:52.30 & $-$35:01:26.5 & 2009-05-21 & $U\!BV\!R$  
& 1800/900/300/240 & 20.4\,$V$\\
          &             &               & 2011-06-29 & grism 4, slit 1.0 
& 7200             & 19.5\,$R$\\
V1310 Sgr & 18:35:01.02 & $-$30:03:35.9 & 2009-05-23 & grism 4, slit 1.5 
& 3600             & 13.5\,$R$\\
V2572 Sgr & 18:31:36.81 & $-$32:35:58.4 & 2009-05-24 & grism 4, slit 1.0 
& 2700             & 17.8\,$R$\\
V728 Sco  & 17:39:05.58 & $-$45:27:14.4 & 2009-05-20 & $U\!BV\!R$ 
& 1800/900/300/180 & 18.5\,$V$\\
          &             &               & 2011-06-29 & grism 4, slit 1.0 
& 1800             & 18.2\,$R$\\
V734 Sco  & 17:45:02.36 & $-$35:38:07.1 & 2009-05-19 & grism 4, slit 1.0 
& 3600             & $<$13.2\,$R$ $^2$\\
\hline\noalign{\smallskip}
\end{tabular}
\\
1) $V$ magnitudes were derived from 
the calibrated EFOSC2 photometry, $R$ magnitudes were estimated by comparing
our acquisition frames with data from the USNO-A2.0 catalogue \citep{monet98-2}
as implemented in the ESO archive. The typical overall uncertainty is 
estimated to $\sim$0.3 mag (for the $V$ magnitudes see Table \ref{ubvr_tab}).\\
2) star is saturated in the acquisition frame
\end{minipage}
\end{table*}

The data were obtained during three observing runs in May 2009, February and 
June 2011, at the ESO-NTT, La Silla, Chile, using EFOSC2 
\citep*{eckertetal89-1}. Only during the
first run the weather conditions allowed for calibrated photometry. The
respective data were taken as $\ge$3 exposures in the $U$, $B$, $V$,
and $R$ passbands. Between each exposure the telescope was moved slightly so 
that pixel deficiencies would average out when combining them to a single 
frame. The individual frames were bias-corrected, but not flatfielded, because 
EFOSC2 flats suffer significantly from a central light concentration problem. 
Subsequently, aperture photometry was performed using IRAF's {\tt phot} 
package. These served as input for the standalone {\tt daomatch} routine 
\citep{stetson92-1} to determine the offsets between the individual frames. 
Correcting for these offsets, the frames were combined to a single one 
using a 3$\sigma$ clipping algorithm for the averaging. Since all photometric 
fields are located in crowded regions the stellar magnitudes were determined 
by fitting the point spread function (PSF). Finally, the magnitudes were 
calibrated using photometric data of standard stars 
\citep{landolt83-1,landolt92-3} that were taken in the same night. 
 
The spectroscopic data were collected using grism 4 and, in one case, 
grism 11, in combination with a 1.0$^{\prime\prime}$ slit. In another 
case, the 1.5$^{\prime\prime}$ slit was employed with grism 4. Similar to
the photometry, the data consisted of three individual spectra that were
later combined prior to the extraction process. The data were corrected
for bias, divided by normalised flats, and afterwards combined and extracted
using IRAF routines. Wavelength calibration was performed using Helium-Argon 
lamps. The resulting typical spectral ranges and resolutions are 3490--7470 
{\AA} at 13 {\AA} (FWHM, i.e.~Full Width at Half Maximum for an arc line) for 
grism 11, and 4050--7440 {\AA} at 11.5 {\AA} (34 {\AA} in case of the 
1.5$^{\prime\prime}$ slit) for grism 4. The spectra were corrected for the
instrumental response function using standard star spectra obtained during
the May 2009 run for grism 4, and the February 2011 run for grism 11. Since
the conditions were not photometric during these nights, these spectra
can not be considered as calibrated in absolute flux, but only regarding the
relative spectral energy distribution (SED).

In order to determine the coordinates reported in Table \ref{obslog_tab},
we have performed an astrometric correction of the combined $R$-band frames 
(or, for those objects without calibrated photometry, of the $R$-band
acquisition frame) using the routines embedded in Starlink's 
GAIA\footnote{\tt http://astro.dur.ac.uk/$\sim$pdraper/gaia/gaia.html} tool
(version 4.4.1) with the UCAC3 \citep{zachariasetal10-1} catalogue. Prior to 
the final fit, saturated stars and ambiguous positions (close visual binaries) 
were manually deleted. The typical RMS of the fit amounted to significantly 
less than 1 pixel, i.e.~less than 0.24$^{\prime\prime}$.

\section{Results}

\begin{figure*}
\includegraphics[width=1.8\columnwidth]{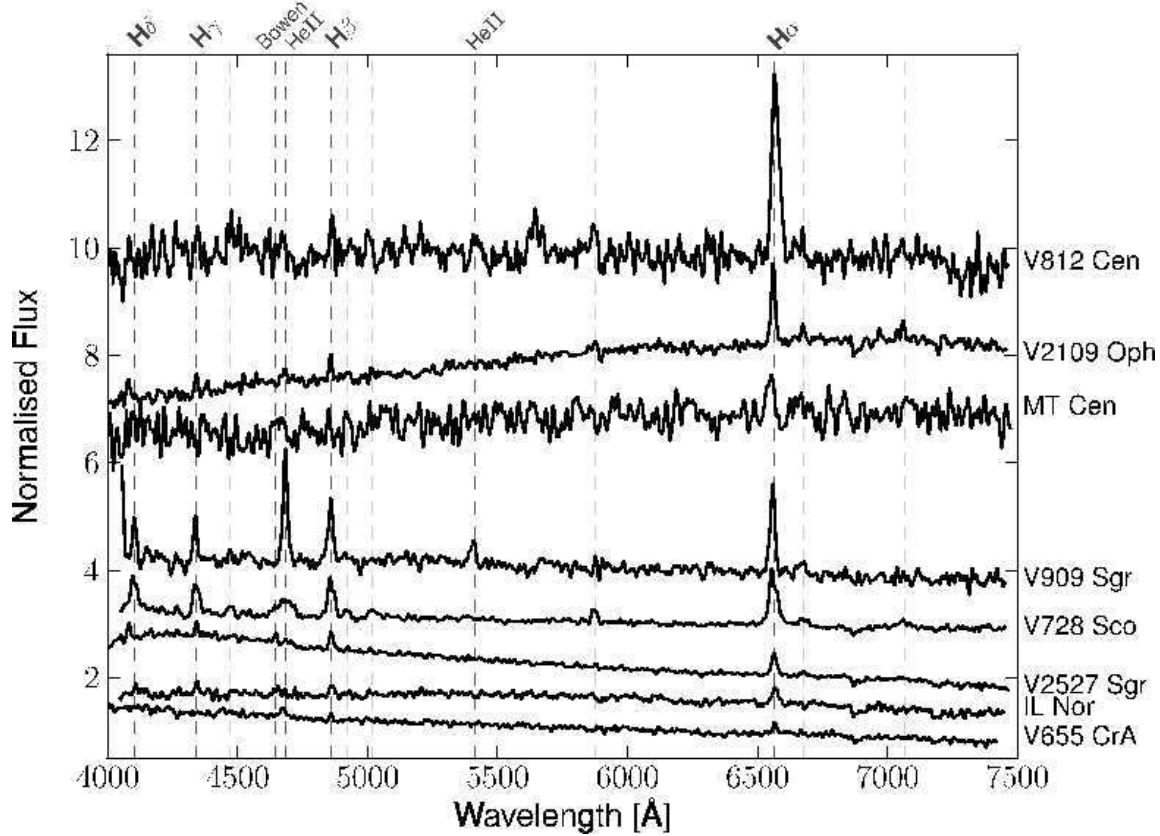}
\caption[]{Spectra of the eight confirmed classical novae. The fluxes have
been normalised by dividing through the mean value, the spectra have
been smoothed by a 3$\times$3 box filter, and they have been shifted 
vertically for clarity. The positions of Balmer, He{\sc II}, and the Bowen
emission features are indicated by the dashed, dark grey, lines and labelled 
at the top of the plot. Unlabelled, light grey, dashed lines mark He{\sc I} 
lines.}
\label{novaspec_fig}
\end{figure*}

\begin{table*}
\begin{minipage}{16cm}
\caption[]{Equivalent widths (in absolute values of angstroms) of the 
identified emission lines for the confirmed post-novae. The errors were 
estimated using a Monte Carlo simulation.}
\label{eqw_tab}
\begin{tabular}{lllllllllllll}
\hline\noalign{\smallskip}
object    &  \multicolumn{4}{c}{Balmer} & \multicolumn{6}{c}{He{\sc I}}
& Bowen/He{\sc}II & He{\sc II} \\
          & 4101  & 4340   & 4861  & 6563    & 4472  & 4922   & 5016   
& 5876$^1$ & 6678 & 7065 & 4645/4686$^2$ & 5412 \\
\hline\noalign{\smallskip}
MT Cen    & --    & --     & 5(2)  & 15(4)   & --    & --     & --     
& --       & --   & --   & 11(4)         & -- \\
V812 Cen  & --    & --     & 15(4) & 120(12) & --    & --     & --     
& 16(7)    & 6(3) & --   & --            & -- \\
V655 CrA  & --    & --     & 1(1)  & 4(1)    & --    & --     & --     
& --       & --   & --   & 3(1)          & -- \\
IL Nor    & 2(1)  & 3(1)   & 4(1)  & 9(1)    & --    & --     & --     
& 1(1)     & --   & --   & 4(1)          & -- \\
V2109 Oph & 20(8) & 14(3)  & 9(2)  & 22(1)   & --    & --     & --     
& 5(1)     & 3(1) & 3(1) & 7(2)          & -- \\
V909 Sgr  & 10(4) & 16(2)  & 24(2) & 39(3)   & 2(1)  & 1(1)   & --     
& 2(1)     & 9(2) & --   & 42(2)         & 10(2) \\
V2572 Sgr & 3(1)  & 2(0.5) & 4(1)  & 13(1)   & --    & 1(0.5) & 1(0.5) 
& 1(0.5)   & 2(1) & --   & 4(1)          & -- \\
V728 Sco  & 12(3) & 16(1)  & 21(1) & 40(1)   & 4(1)  & 3(2)   & 6(1)   
& 5(1)     & 4(1) & 8(1) & 21(1)         & -- \\
\hline\noalign{\smallskip}
\end{tabular}
\\
1) this line is mostly distorted by the adjacent Na{\sc I} absorption.\\
2) our resolution does not permit to properly resolve these components.
\end{minipage}
\end{table*}

\begin{table}
\caption[]{Results of the $U\!BV\!R$ photometry.}
\label{ubvr_tab}
\begin{tabular}{lllll}
\hline\noalign{\smallskip}
object & $V$ & $U\!-\!B$ & $B\!-\!V$ & $V\!-\!R$ \\
\hline\noalign{\smallskip}
MT Cen   & 19.79(12) & $-$0.04(03) & 1.10(02) & 0.47(02) \\
V812 Cen & 21.32(15) & $-$0.57(07) & 0.57(06) & 0.54(05) \\
IL Nor   & 19.03(14) & $-$0.69(01) & 0.29(02) & 0.34(02) \\
V909 Sgr & 20.39(14) & $-$0.49(02) & 0.32(02) & 0.20(02) \\
V728 Sco & 18.47(13) & $-$0.83(01) & 0.42(01) & 0.60(01) \\
\hline\noalign{\smallskip}
\end{tabular}
\end{table}

\begin{figure}
\includegraphics[width=\columnwidth]{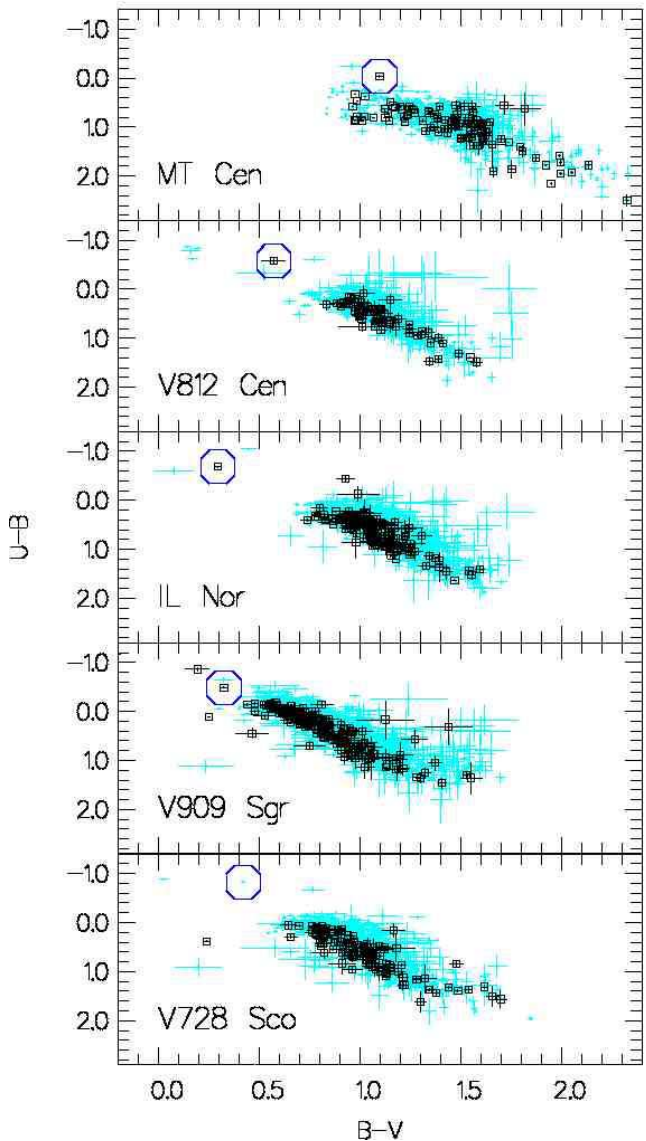}
\caption[]{Colour-colour diagram ($U\!-\!B$ vs $B\!-\!V$) for the fields of
MT Cen, V812 Cen, IL Nor, V909 Sgr, and V728 Sco. The black squares indicate 
objects within the central 300$\times$300 pixels 
($\sim$1.2$\times$1.2$^\prime$). The post-nova is marked with a circle.}
\label{ubv_fig}
\end{figure}

\subsection{MT Cen = Nova Cen 1931}
\label{mtcen_sec}

\begin{table}
\caption[]{Galactic coordinates and corresponding interstellar extinction 
\citep*[taken from NASA's IRSA web interface]{schlegeletal98-2}.}
\label{extinc_tab}
\begin{tabular}{llll}
\hline\noalign{\smallskip}
object & $l$ & $b$ & $E(B\!-\!V)$ \\
       &     &     & [mag]\\
\hline\noalign{\smallskip}
MT Cen    & 294.7 & $+$1.2  & 1.60 \\
V812 Cen  & 305.9 & $+$5.1  & 0.50 \\
V655 CrA  & 356.9 & $-$11.1 & 0.12 \\
IL Nor    & 326.8 & $+$4.8  & 0.58 \\
V2109 Oph & 1.1   & $+$6.3  & 0.92 \\
V909 Sgr  & 358.8 & $-$10.4 & 0.11 \\
V1310 Sgr & 4.2   & $-$10.0 & 0.12 \\
V2572 Sgr & 1.5   & $-$10.4 & 0.15 \\
V728 Sco  & 345.2 & $-$7.6  & 0.37 \\
V734 Sco  & 354.2 & $-$3.4  & 0.94 \\
\hline\noalign{\smallskip}
\end{tabular}
\end{table}

\begin{figure}
\includegraphics[width=\columnwidth]{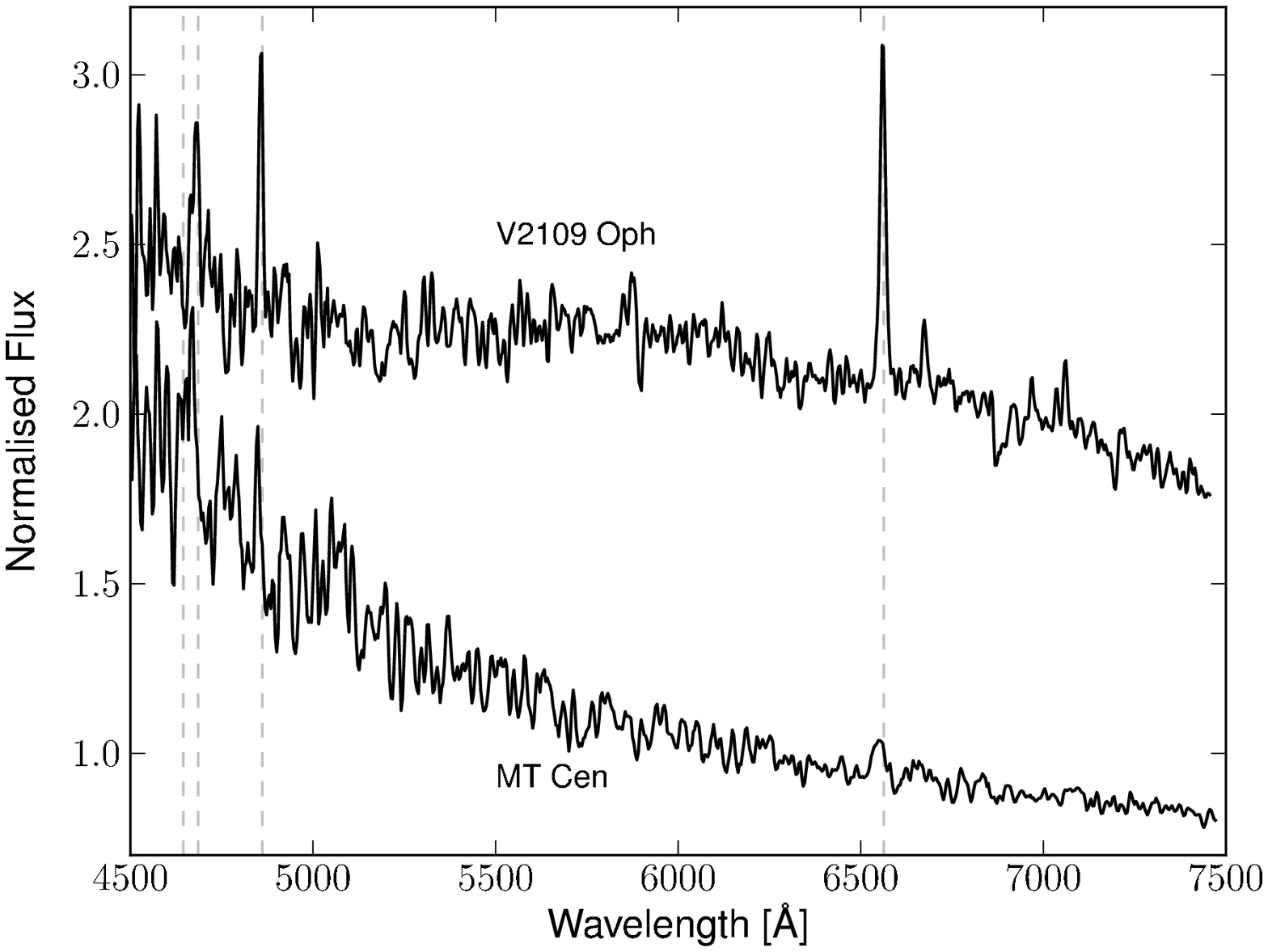}
\caption[]{Dereddened spectra of MT Cen and V2109 Oph, the two novae that 
are the most strongly affected by extinction (Table \ref{extinc_tab}).
Like in Fig.~\ref{novaspec_fig}, the spectra have been smoothed with a
3$\times$3 box filter, normalised by dividing through the mean value, and
displaced vertically for clarity. The dashed vertical lines mark (from
left to right) the positions of the Bowen blend, He{\sc II} 4686, H$\beta$
and H$\alpha$.}
\label{deredspec_fig}
\end{figure}

MT Cen was classified as a fast nova ($t_3 \sim 10~\mathrm{d}$) that reached
a photographic brightness of 8.4 mag on May 12, 1931 \citep{duerbeck84-5}.
Due to its faintness already the later stages of the eruption are only
sparsely covered, and first attempts to observe the post-nova or its shell
were unsuccessful \citep[][respectively]{munari+zwitter98-1,gill+obrien98-1},
Later, \citet{woudt+warner02-5} identified a likely candidate for the
post-nova by detecting flickering-like variability in their high-speed 
photometry.

Our colour-colour diagram (Fig.~\ref{ubv_fig}, top) points out this same 
object as having colours different from the majority of the stars in the
field. This becomes more pronounced if only the central part of the
CCD image is considered (black squares in Fig.~\ref{ubv_fig}). Our spectrum
of the candidate has very low S/N (Fig.~\ref{novaspec_fig}), but the detection 
of weak H$\alpha$ emission confirms the post-nova. In agreement with
the not particularly blue colour derived photometrically (Table 
\ref{ubvr_tab}), the continuum flux increases toward longer wavelengths.
One possibility is that the late-type donor contributes significantly to
the optical flux, which would indicate either a low-mass-transfer system, or
a bright donor, the latter implying a long orbital period. However,
this part of the sky suffers from considerable reddening (Table 
\ref{extinc_tab}). In order to examine the influence of the extinction
we have dereddened the spectrum using the corresponding IRAF task, which is
based on the relations derived by \citet*{cardellietal89-1}. Since the
absolute extinction $A(V)$ is not known, we have used the standard value
for the ratio $R(V) = A(V)/E(B\!-\!V) = 3.1$. The resulting dereddened
spectrum is shown in Fig.~\ref{deredspec_fig}. It proves that the red slope of
the continuum was entirely due to interstellar reddening, which even masked 
the presence of the emission components of H$\beta$ and the Bowen blend. The
corrected spectrum shows an SED that is typical for a high mass-transfer
system.

\subsection{V812 Cen = Nova Cen 1973}

This object was a late discovery, reported five years after its eruption
by \citet*{macconnelletal78-1} who detected a typical spectrum of a nova in its
nebular phase on objective prism plates. The authors determined the continuum
brightness at this stage to $V = 11~\mathrm{mag}$. \citet{duerbeck87-1}
identified a candidate for the post-nova, but attempts to confirm it
spectroscopically \citep{zwitter+munari96-1} or by detecting the nova shell
\citep{downes+duerbeck00-1} remained unsuccessful. 

In our photometry we find this candidate as a blue object about 10 mag
fainter than the nova at the time of its detection (Fig.~\ref{ubv_fig},
Table \ref{ubvr_tab}). The spectroscopy shows a flat continuum with strong 
H$\alpha$ emission, much weaker H$\beta$, as well as a couple of He{\sc I}
lines (Fig.~\ref{novaspec_fig}, Table \ref{eqw_tab}). The Balmer decrement is 
much stronger than usual in CVs \citep[e.g.,][]{williams83-1}. 
Since even 
after the dereddening the continuum slope is not particularly steep
(Table \ref{erup_tab}), it does not appear that either of the underlying 
stellar components of the binary is affecting the strength of the emission 
lines. Rather it seems more likely that an additional, non-stellar, source is 
contributing to the H$\alpha$ emission. Although \citet{downes+duerbeck00-1} 
did not detect a nova shell, such a shell still represents the most promising 
candidate for the additional H$\alpha$ source.

This is also the youngest nova in our sample, although three other systems 
with ``normal" values are not far behind (Table \ref{erup_tab}). 
\citet*{ringwaldetal96-3} found that the contribution by the nebula does not 
only depend on the time that has passed since the eruption, but also on the 
speed class, with the nebular contribution being more important for slower 
novae. Unfortunately there is no corresponding information available for V812 
Cen, 
and we can thus only speculate that this nova has not been a 
particularly fast one.
A similar, even slightly stronger, Balmer decrement was found by 
\citet{ringwaldetal96-3} for FH Ser ($t_3 = 62~\mathrm{d}$) in a spectrum 
taken 20 years after the eruption.

\subsection{V655 CrA = Nova CrA 1967}

Similar to V812 Cen, this nova was discovered some time after its original
eruption on objective prism plates \citep{sanduleak69-4}. The author 
determined the magnitude at the time of the discovery to 13 mag. Based on
the characteristics of the spectrum he estimated the maximum brightness to
8 mag. He furthermore identifies a likely progenitor with a red magnitude
of 17 mag, a value that was later corrected to $J = 17.6~\mathrm{mag}$ by
\citet{duerbeck87-1}. However, the pre-nova candidate turned out to be
a close visual binary, and the correct identification remained unclear.
\citet{duerbeck+seitter87-1} took a spectrum, but it was of too low quality
to even unambiguously detect the presence of emission lines. Finally,
\citet{woudt+warner03-4} identified a likely candidate with a mean magnitude
of $V = 17.6$ based on flickering-type variability detected in their 
high-speed photometry.

Our spectrum (Fig.~\ref{novaspec_fig}) confirms this object as the
post-nova. It shows a steep blue continuum with comparatively narrow
emission of H$\alpha$ and H$\beta$, as well as a broader Bowen/He{\sc II}
4686 component (Table \ref{eqw_tab}). This system likely still sustains a 
high mass-transfer rate.

\subsection{IL Nor = Nova Nor 1893}

This is the second oldest nova in this sample. It was discovered by M.~Fleming
on objective prism plates \citep[as reported by][]{pickering93-6}.
Investigations of earlier photographic plates yielded a maximum magnitude of
7 \citep{pickering93-7}. \citet{duerbeck87-1} classifies it as a moderately 
fast nova. The post nova is a member of a close visual triple system that
for some time could not be resolved, although \citet{duerbeck+seitter87-1}
reported a spectrum showing Balmer emission lines. An attempt by
\citet{gill+obrien98-1} to image the nova shell was unsuccessful. In a
recent paper, \citet{woudt+warner10-1} identified the post-nova in their
high-speed photometry. The authors present two short light curves about
one year apart (March 2003 and February 2004). Apart from some shorter
term variability the second run showed the object on average 0.5 mag fainter 
than the first one.

Our photometry (Table \ref{ubvr_tab}, Fig.~\ref{ubv_fig}) reveals this 
candidate as a very blue object with $V = 19.0~\mathrm{mag}$, at the same 
brightness as in the second photometric run presented by 
\citet{woudt+warner10-1}. 
The spectrum consists of a blue continuum, with only 
weak Balmer and He{\sc I} emission. A Bowen/He{\sc II} component is also
clearly visible. These characteristics suggest that more than a hundred years
after its eruption this system still drives a comparatively high mass-transfer 
rate, which is in agreement with the results by \citet{ringwaldetal96-3} that
generally the spectroscopic properties of novae after the initial decline 
phase (20--30 yr) change very little over the next decades.

\subsection{V2109 Oph = Nova Oph 1969}

Another late discovery, this nova was reported eight years after its
eruption when \citet{macconnell77-2} detected an object with a red magnitude
of 10.8 and an emission line spectrum on archival objective prism plates. 
A later study on Sonneberg plates by \citet{wenzel92-4} revealed that the
discovery plate missed maximum brightness by at least twelve days, when the
object presented a blue magnitude of 8.9. \citet{duerbeck87-1} identified
a likely candidate for the post-nova, that was later observed by 
\citet{szkody94-2} as a comparatively red star with $V = 19.4~\mathrm{mag}$,
$B\!-\!V > 1.4$ and $V\!-\!R = 1.2$.

Our spectroscopy finds the object at $R \sim 19.7~\mathrm{mag}$, and thus
much fainter than during the \citet{szkody94-2} observations that took
place 20 years after the eruption ($R = 18.2~\mathrm{mag}$). 
The spectrum shows the red continuum that could be expected from Szkody's
photometry. Like in the case of MT Cen (Sec.~\ref{mtcen_sec}) this is in 
large part due to the increased extinction in this area of the sky 
(Table \ref{extinc_tab}). Correcting for this
effect yields a considerably bluer spectrum (Fig.~\ref{deredspec_fig}),
although the SED is still rather unusual for a post-nova (see discussion
in Sec.~\ref{disc_sec}). We 
furthermore find moderately strong emission lines of the Balmer and He{\sc I} 
series, as well as He{\sc II} 4686 (Fig.~\ref{novaspec_fig}, 
Table \ref{eqw_tab}). Contrary to most of the other systems in this sample, 
the Bowen blend cannot be detected in V2109 Oph, indicating a comparatively 
low mass-transfer rate.

\subsection{V909 Sgr = Nova Sgr 1941}

\citet{duerbeck87-1} summarises the discovery history of this nova that
erupted in 1941, reached a photographic maximum brightness of 6.8 mag,
and showed a very fast decline rate of $t_3 = 7~\mathrm{d}$. 
\citet{diaz+bruch97-1} report the post-nova as being an eclipsing system
with an orbital period of 3.36 h, but emphasise the need for confirmation.
Since they do not provide a finding chart, the correct identification of
the post-nova remained ambiguous.

In our colour-colour diagram (Fig.~\ref{ubv_fig}) we detect a blue object
very close to the coordinates listed in the \cite{downesetal05-1} catalogue.
The spectroscopy shows a blue continuum superposed with hydrogen and helium
emission lines (Fig.~\ref{novaspec_fig}). Certainly the most remarkable feature 
is the strength of He{\sc II} in the system, evidenced in the presence of the 
$\lambda$5412~{\AA} line, and the $\lambda$4686~{\AA} line rivalling H$\alpha$
in strength, while the Bowen blend appears absent (Table \ref{eqw_tab}). The 
spectrum -- at least at the current resolution and S/N -- bears a strong 
resemblance to that of the old nova and asynchronous polar V1500 Cyg 
\citep[e.g.,][]{ringwaldetal96-3}, whose 3.35~h orbital period 
\citep*{patterson79-7,semeniuketal95-2} coincidentally is very close to the 
one reported by \citet{diaz+bruch97-1} for V909 Sgr.

\subsection{V1310 Sgr = Nova Sgr 1935}

\begin{figure}
\includegraphics[width=\columnwidth]{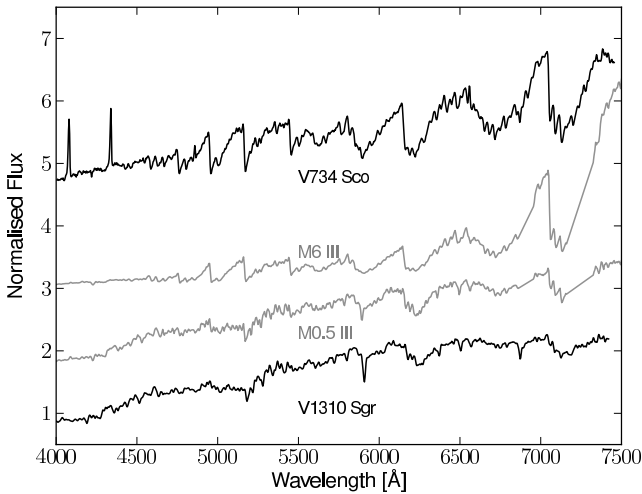}
\caption[]{Spectra of the two suspected Mira stars, V1310 Sgr and V734 Sco.
The data have been dereddened as described in Sec.~\ref{mtcen_sec}.
For comparison, two spectra from the \citet{silva+cornell92-1}
catalogue are overplotted (light grey). Note that the wavelength regions
of atmospheric absorption, 6865--6950 {\AA} and 7165--7320 {\AA}, in the
latter data are undefined, and in the plot linearly interpolated.}
\label{miraspec_fig}
\end{figure}

This star was flagged as a nova by A.~D.~Fokker in 1951 \citep[as reported by]%
[]{duerbeck87-1} based on photometric variability. Duerbeck classifies it as
a slow nova with a ``steep rise and [an] extremely slow, fairly smooth 
decline". The time to drop by 3 magnitudes from the photographic maximum 
brightness of 11.7 mag was measured as $t_3 = 390~\mathrm{d}$. There is no 
spectroscopic information available, neither near maximum brightness, nor at 
minimum. \citet{downesetal05-1} mark a fairly bright star \citep[the 2MASS 
catalogue gives $R = 13.2~\mathrm{mag}$;][]{cutrietal03-3} as the post-nova, 
but remark that the object is possibly a Mira variable instead of a nova.

Our spectroscopy finds the star at $R \sim 13.5~\mathrm{mag}$. The spectrum
(Fig.~\ref{miraspec_fig}) looks like a late K or early M star. 
No emission lines are observed. 
We used the TiO5 index defined and calibrated by \citet*{reidetal95-3} to 
calculate a spectral type of M0$\pm$0.5. This is probably a Mira type star.

This poses the question if V1310 Sgr has to be qualified as a misidentification 
(i.e.~the post-nova is some other object in the field) or as a 
misclassification (i.e.~the original discovery report mistook the Mira
variability for a nova eruption). Since both the shape and the time scale of 
the photometric variability as well as the observed difference in magnitude 
($\sim$2.5 mag) are consistent with a Mira type light curve \citep[e.g.,][]%
{lebzelter11-1}, we favour the latter possibility, and suggest to remove 
V1310 Sgr from the list of potential classical novae.

\subsection{V2572 Sgr = Nova Sgr 1969}

The nova was discovered due to its photometric variability that reached a 
maximum photographic brightness of 6.5 mag as reported by \citet{bateson69-7}.
\citet{duerbeck87-1} classifies it as a moderately fast nova with 
$t_3 = 44~\mathrm{d}$. The eruption light curve is given by \citet{knight72-3},
who furthermore reports a ``fairly constant" brightness of 
$V \sim 13~\mathrm{mag}$ for the pre-nova in the years 1957--1968, as well as 
for the post-nova in 1970--1972. Consistent with these findings, 
\citet{radiman+hidajat75-3} present $V \sim 12.5 \pm 0.5~\mathrm{mag}$ for the 
pre-nova in the years 1966--1967. However, in the same time range the authors 
also find an unusually red and strongly variable colour for this object to 
$B\!-\!V = 0.8-2.0~\mathrm{mag}$. Since modern finding charts 
\citep{downesetal05-1} show a fairly crowded region of the sky at the given
position, it is thus likely that above values for the pre- and post-nova
do not represent the nova itself, but rather the combined brightness of the 
nova and its close visual companions.

Supporting this interpretation we find the nova to have a steep blue continuum,
and a much fainter brightness than reported of $R \sim 17.8~\mathrm{mag}$.
Emission lines of the Balmer and He{\sc I} series are clearly discernible,
albeit comparatively weak. A Bowen/He{\sc II} component is also present.
Overall, the spectroscopic characteristics give the impression of a
high mass-transfer system. On our acquisition frame (Fig.~\ref{fc_novae02_fig})
we note that the vicinity of the post-nova appears ``smudgy", which could
indicate the presence of a nova shell.

\subsection{V728 Sco = Nova Sco 1862 = Nova Ara 1862}

\begin{figure}
\includegraphics[angle=270,width=\columnwidth]{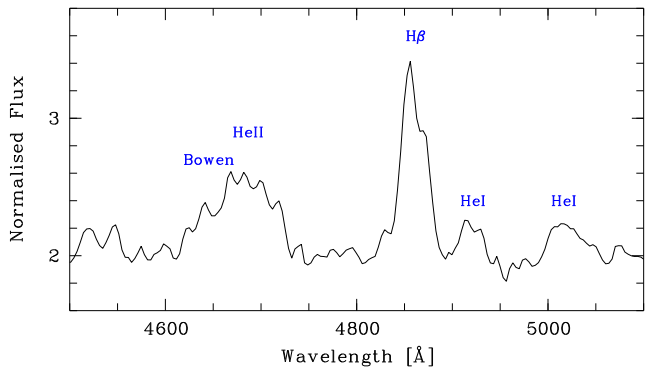}
\caption[]{Close-up of the spectrum of V728 Sco. The data have been smoothed
by a 3$\times$3 box filter.}
\label{v728scosp_fig}
\end{figure}

Close to the border between constellations Scorpius and Ara (which is why
it was originally assigned to the latter) a bright star of 5$^\mathrm{th}$
magnitude was reported by \citet{tebbutt78-3} to have been observed visually
on October 5--9, 1862. Only four days later he found it to have declined to
below 11$^\mathrm{th}$ mag. \citet{duerbeck87-1} identified two faint
candidates ($j \sim 20-21~\mathrm{mag}$) for the post-nova based on Tebbutt's 
coordinates. \citet{diaz+steiner91-1} list the object as a potential
magnetic nova due to its high eruption amplitude and fast decline. However,
\citet{schmidtobreicketal02-1} note that this candidate presents colours
that are more consistent with a main-sequence star than a CV. Based on their
colour-colour diagram they instead suggest two new candidates that are within
1$^\prime$ of the original coordinates.

Our present photometric observations cover a larger area on the sky. The
corresponding colour-colour diagram is presented in Fig.~\ref{ubv_fig}. The 
two candidates from \citet{schmidtobreicketal02-1} can be found as the
bluest object in the field with $U\!-\!B = -0.88$, $B\!-\!V = 0.02$, and
as the black square at $U\!-\!B = 0.38$, $B\!-\!V = 0.24$. Spectroscopic
observations of the latter, which are not presented here in detail, showed
it to be a blue star with Balmer absorption lines. The other candidate has
not been observed spectroscopically. 
Instead we find the post-nova to be
an object slightly more than 2$^\prime$ northwest of the position given
in the \citet{downesetal05-1} catalogue (see Table \ref{obslog_tab} and
Fig.~\ref{fc_novae02_fig} for the corrected position).
The spectroscopy 
reveals a blue continuum and (for a nova) comparatively strong Balmer and 
He{\sc I} 
emission lines (Fig.\,\ref{novaspec_fig}, Table \ref{eqw_tab}). This system is 
the one in our sample that looks most like a dwarf nova 
\citep[e.g.,][]{williams83-1} but for the blue slope and the presence of a 
broad Bowen/He{\sc II} component. The broad and structured Balmer emission 
lines (Fig.~\ref{v728scosp_fig}) make it an interesting object for further 
studies, because they suggest a high-inclination, possibly eclipsing, CV.

\subsection{V734 Sco}

As summarised by \citet{duerbeck87-1} this object was reported as a possible 
long-period variable or nova by L.~Plaut who discovered it to be brighter than
15$^\mathrm{th}$ mag for 10 days on photographic plates from 1937. The star
is bright in the infrared: $J = 8.7~\mathrm{mag}$, $H = 7.6~\mathrm{mag}$,
$K = 6.9~\mathrm{mag}$ \citep{cutrietal03-3}, and listed as a probable Mira
variable in the \citet{downesetal05-1} catalogue.

Our spectrum in Fig.~\ref{miraspec_fig} shows TiO bands of a cool star and 
Balmer emission lines. The strongest emission is observed in H$\delta$ 
(equivalent width $W_\lambda$ = $-$37 {\AA}), followed by 
H$\gamma$ ($W_\lambda$ = $-$22 {\AA}) and probably 
H$\beta$. This unusual Balmer decrement is typical for oxygen-rich (M-type) 
Mira stars \citep{castelazetal00-2}. 
We used the TiO5 index defined by \citet{reidetal95-3} and calibrated by 
\citet{cruz+reid02-1} to calculate a spectral type of M8.7$\pm$0.5. The 
overall shape around 6830 {\AA} indicates a giant rather than a dwarf star, 
consistent with a Mira classification. 

Like in the case of V1310 Sgr we therefore conclude that V734 Sco is not
a classical nova and should not be listed as such.

\section{Discussion}
\label{disc_sec}

\begin{table*}
\begin{minipage}{13cm}
\caption[]{Properties of the confirmed post-novae. See text for details.}
\label{erup_tab}
\begin{tabular}{llllllll}
\hline\noalign{\smallskip}
object & $m_\mathrm{max}$$^1$ & $m_\mathrm{min}$ & $\Delta m$ & $\Delta t$ 
& $t_3$ & $\alpha$ & $F\!W\!H\!M(\mathrm{H}\beta$) \\
       & [mag] & [mag] & [mag] & [yr] & [d] & & [{\AA}]\\
\hline\noalign{\smallskip}
MT Cen    & 8.4\,$p$     & 19.8\,$V$ & 11.4    & 78  & $\sim$10 
& 4.45(07) & 14 \\
V812 Cen  & $<$11.0\,$V$ & 21.3\,$V$ & $>$10.3 & 36  & --       
& 1.75(11) & 16 \\
V655 CrA  & 8.0\,$p$     & 17.7\,$R$ & 9.7     & 42  & --       
& 1.44(02) & 10 \\
IL Nor    & 7.0\,$p$     & 19.0\,$V$ & 12.0    & 116 & 108      
& 2.53(03) & 21 \\
V2109 Oph & 8.9\,$b$     & 19.7\,$R$ & 10.8    & 40  & --       
& $-$0.21(03) / 1.82(08) $^2$ & 13 \\
V909 Sgr  & 6.8\,$p$     & 20.4\,$V$ & 13.6    & 68  & 7        
& 1.41(04) & 20 \\
V2572 Sgr & 6.5\,$p$     & 17.8\,$R$ & 11.3    & 40  & 44       
& 2.19(02) & 16 \\
V728 Sco  & 5.0\,$v$     & 18.5\,$V$ & 13.5    & 147 & $<$9     
& 1.85(03) & 30 \\
\hline\noalign{\smallskip}
\end{tabular}
\\
1) $p$: photographic, $V$: $V$-band, $b$: blue, $v$: visual\\
2) first value for $\lambda < 5900~\mathrm{\AA}$, second value for $\lambda 
\ge 5900~\mathrm{\AA}$
\end{minipage}
\end{table*}

While the present work is mainly aimed at increasing the sample of confirmed
old novae for later studies, and the data quality is mostly not suited for a
thorough analysis, it still allows to remark on one or the other detail.

In Table \ref{erup_tab} we list several photometric and spectroscopic
properties of the confirmed post-novae. We have used the previously reported 
maximum brightness $m_\mathrm{max}$ (in column 2) and the current brightness 
as derived from our observations $m_\mathrm{min}$ (column 3) to derive the
eruption amplitude $\Delta m = m_\mathrm{max}-m_\mathrm{min}$ (column 4). This
calculation ignores brightness differences between filters. Taking into account
that the listed magnitudes are close to the visual range, and that colour
differences within the $B-R$ range rarely exceed 1.0 mag (see 
Table \ref{ubvr_tab}), we can estimate a typical uncertainty of $\sim$0.5 mag.
Column 5 presents the ''age" of the post-nova, i.e.~the time that has passed 
from the eruption to the current observations, and column 6 summarises the
time $t_3$ in which the nova has declined by 3 magnitudes as taken from 
\citet{duerbeck87-1}. 

In order to examine the SED of the confirmed novae, we have fitted a
power law $F \propto \lambda^{-\alpha}$ to the continuum of the dereddened 
spectra. We restricted the fit to wavelengths $5000 - 7200~\mathrm{\AA}$ 
because the continuum bluewards of $5000~\mathrm{\AA}$ in many systems is less
well defined due to the presence of several emission lines, and also
the noise in general increases towards the blue end of the spectrum.
The such derived values for the power-law exponent are given in column 7
of Table \ref{erup_tab}. Note that the corresponding uncertainty listed
there corresponds to the standard deviation of the fit, and does not take
into account additional uncertainties concerning the instrumental response
function and the dereddening process. Finally, in column 8 we list the
Full Width at Half Maximum of a Gaussian fit to the H$\beta$ line. We have
used this line instead of the stronger and usually better defined H$\alpha$
line in order to avoid potential contamination of the shell material.

Starting with the SED, we find that $\alpha$ for most systems falls well 
below the value of 2.33 for a steady-state accretion disc 
\citep{lynden-bell69-2}. This is in agreement with \citet{wade88-1} who
places nova-like CVs in this range, but differs from the results of 
\citet{ringwaldetal96-3} who derive an average value of $\alpha = 2.68$ with 
a standard deviation of $0.82$ for their sample of dereddened post-novae.
Within our sample, MT Cen stands out by presenting with $\alpha = 4.45$ 
by far the largest value. In principle this could mean that this system
is dominated by a still hot dwarf, but considering that it was observed
78 years after the eruption, this appears unlikely. We furthermore note
that the largest correction for interstellar extinction had to be used
for MT Cen. This might be coincidental, but there is also the possibility
that the uncertainties in the dereddening process are the reason behind
MT Cen's exalted position, and perhaps also behind the difference between our
results and \citet{ringwaldetal96-3}. Another system with an SED 
significantly different from the others in our sample is V2109 Oph, whose
continuum cannot be fitted by a single power-law, but instead follows
different laws for the range redwards of 5900~{\AA} ($\alpha = 1.82$)
and for the range 5000--5900~{\AA} ($\alpha = -0.21$). A similar phenomenon
has been observed by \citet{schmidtobreicketal05-2} for the old novae
V630 Sgr and V842 Cen, of which the former is more similar to our case,
since it also presents the less steep continuum slope for the blue part of its 
SED. The authors interpret their findings as the signature of a disrupted 
inner accretion disc, and, encouraged by the strong He{\sc II} emission in 
the system, suggest that V630 Sgr is a magnetic CV. In V2109 Oph, however, 
the He{\sc II} line is not particularly strong for an old nova, and with the 
present data there is no reason to suspect magnetic accretion. Instead, 
Fig.~\ref{deredspec_fig} shows that the continuum slope bluewards of 
5000~{\AA} again increases and appears similar to that redwards of 5900~{\AA}. 
Such ``bumpy" continuum could in principle be the signature of an SED that
is dominated by the stellar components instead of the accretion disc. Further
evidence for this, e.g.~in the form of stellar absorption lines, however is
missing. Without further data it therefore remains unclear if the shape
of the continuum in V2109 Oph is intrinsic or an artifact due to a problem
in the extraction or dereddening process.

Looking at the eruption amplitudes we find that two systems, V909 Sgr and
V728 Sco, match the criterion of \citet{schmidtobreicketal04-4} for a
Tremendous Outburst Nova (TON), $\Delta m > 13~\mathrm{mag}$. Based on the
assumption that the absolute eruption magnitude is very similar for all
novae the authors \citep[see also ][]{schmidtobreicketal05-2} suggest that a 
large eruption amplitude indicates a faint post-nova, either because it is 
seen at high inclination \citep{warner87-4}, or because it is intrinsically
faint due to a low mass-transfer rate. V728 Sco could fit into the former
category. It is the only system in our sample whose emission lines are broad
enough to be well-resolved (last column in Table \ref{erup_tab}), and
additionally these lines show a distinctive structure 
(Fig.~\ref{v728scosp_fig}). Following \citet{diaz+steiner91-1} a large
eruption amplitude and a fast decline can also indicate a magnetic nova,
since discless systems, or those with a disrupted disc, should be
intrinsically fainter than disc CVs. Since V909 Sgr additionally presents a
particularly strong He{\sc II} emission, it is a good candidate for this 
category.

\section{Summary}

We have conducted a study on ten candidate old novae. Among them we have
found two probable Mira stars whose variability likely was confused with a nova
eruption, and we propose to delete these stars, V1310 Sgr and V734 Sco,
from the list of potential novae. We have furthermore spectroscopically
confirmed seven previously selected candidate post-novae, as well as
recovered one system, V728 Sco, that was found to be $\sim$2$^\prime$ away
from the reported coordinates.

Within the confirmed old novae we find a group of four objects that --
after correction for interstellar reddening -- share very similar spectral
properties: MT Cen, V655 CrA, IL Nor, and V2572 Sgr, all have the blue
continuum and the weak emission lines that are typical for high mass-transfer
systems. The remaining four novae instead present at least one pecularity
that distinguishes them from the other systems in our sample. V2109 Oph
presents a ''bumpy" continuum for which with the current data we do not find 
an explanation, and V909 Sgr shows several trademarks of a magnetic systems.
V812 Cen has an unususally strong Balmer decrement, and one can assume that 
the ejected material still contributes to the H$\alpha$ emission. 
Last, not least, V728 Sco is the only system in our sample where we find 
convincing evidence that it has a comparatively high inclination, and it 
should therefore be possible to derive its orbital period with time-series 
photometry.

\section*{Acknowledgments}
{\em In fond memory of Hilmar W.~Duerbeck (1948--2012)}\\[0.1cm]
We would like to thank the referee, Mike Shara, for giving us the thumbs up.

This research was supported by FONDECYT Regular grant 1120338 (CT and NV).
AE acknowledges support by the Spanish Plan Nacional de Astrononom\'{\i}a y 
Astrof\'{\i}sica under grant AYA2011-29517-C03-01. REM acknowledges support by  
the Chilean Center for Astrophysics FONDAP 15010003 and from the BASAL
Centro de Astrofisica y Tecnologias Afines (CATA) PFB--06/2007.

We gratefully acknowledge ample use of the SIMBAD database, 
operated at CDS, Strasbourg, France, and of NASA's Astrophysics Data System 
Bibliographic Services. The Digitized Sky Surveys were produced at
the Space Telescope Science Institute under U.S. Government grant NAG W-2166, 
based on photographic data obtained using the Oschin Schmidt Telescope on 
Palomar Mountain and the UK Schmidt Telescope. We have furthermore made use 
of the NASA/ IPAC Infrared Science Archive, which is operated by the Jet 
Propulsion Laboratory, California Institute of Technology, under contract with 
the National Aeronautics and Space Administration. IRAF is distributed by the 
National Optical Astronomy Observatories.

\appendix

\section{Finding charts}

Here we present finding charts for the eight confirmed old novae. Where
available the combined EFOSC2 $R$ band photometric images were used, 
otherwise the EFOSC2 acquisition frames were employed. The two discarded 
nova candidates, V1310 Sgr and V734 Sco, can be unambiguously identified 
on finding charts in the \citet{downesetal05-1} catalogue.

\begin{figure*}
\includegraphics[width=2.\columnwidth]{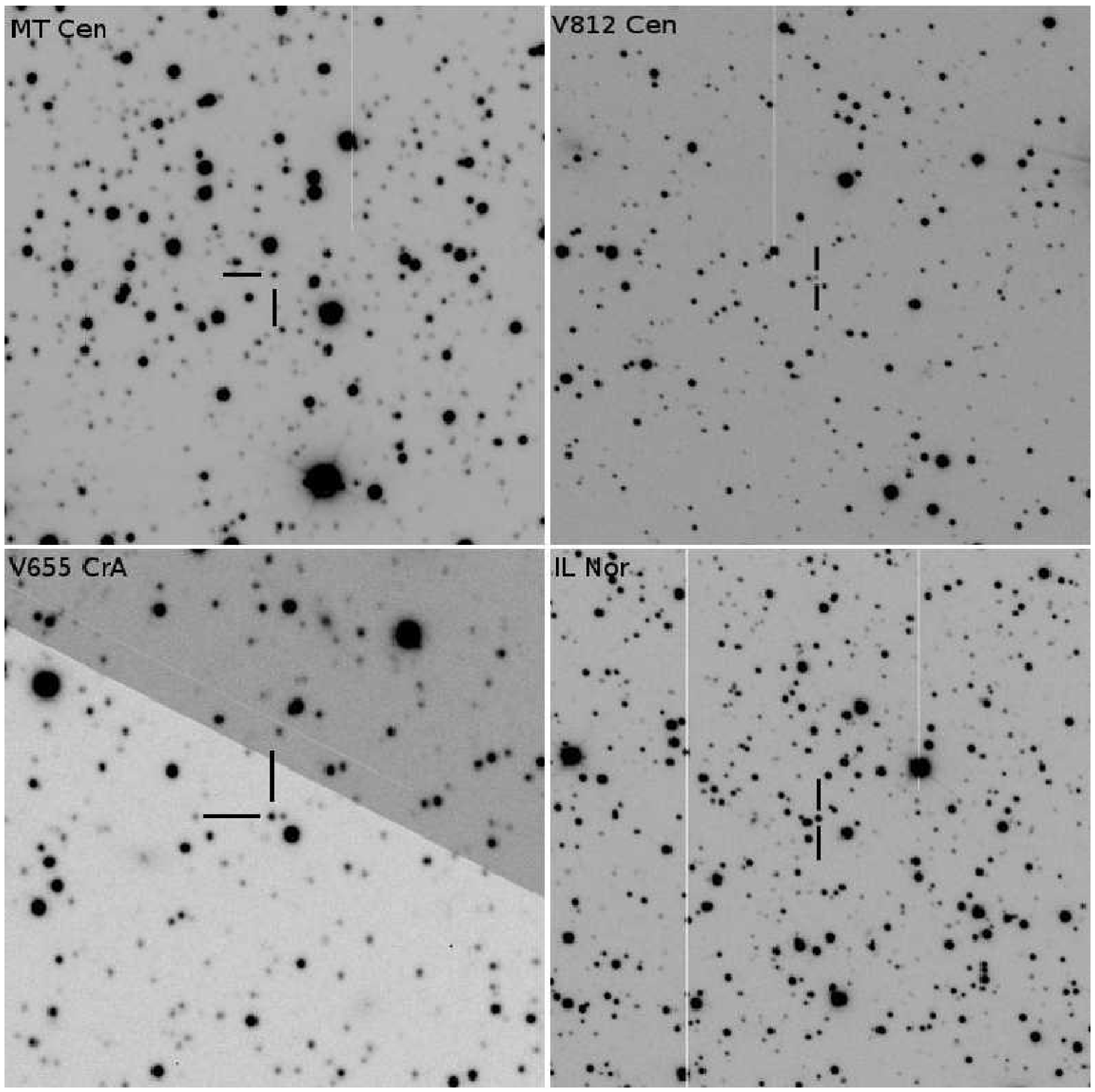}
\caption[]{Finding charts for the confirmed old novae MT Cen, V812 Cen, 
V655 CrA, and IL Nor. The size of a chart is 
1.5$^\prime$ $\times$ 1.5$^\prime$, and the orientation is such that North 
is up and East is to the left. The images were taken in the $R$ band.}
\label{fc_novae01_fig}
\end{figure*}

\begin{figure*}
\includegraphics[width=2.\columnwidth]{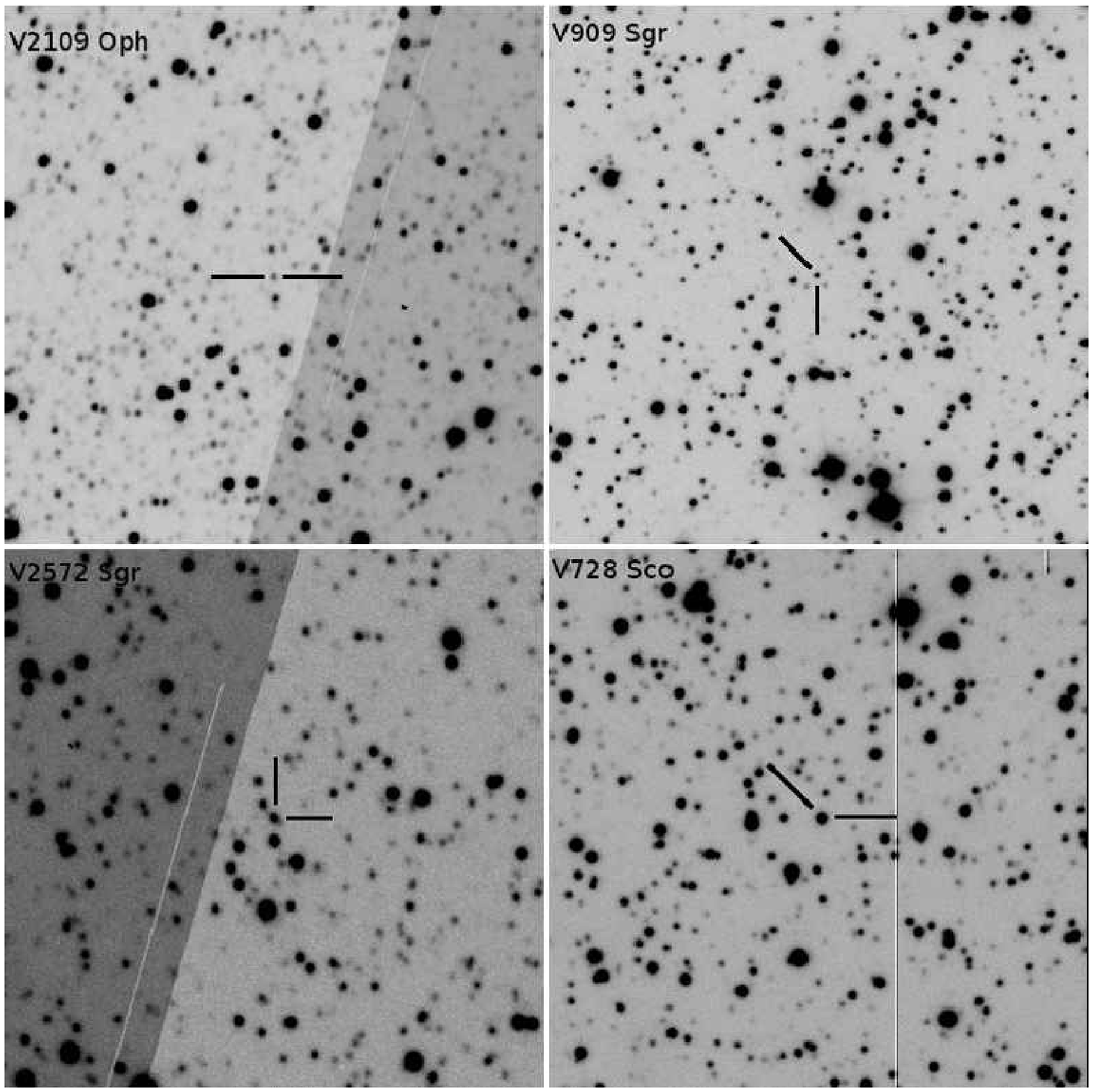}
\caption[]{Finding charts for the confirmed old novae V2109 Oph, V909 Sgr, 
V2572 Sgr, and V728 Sco. The size of a chart is 
1.5$^\prime$ $\times$ 1.5$^\prime$, and the orientation is such that North 
is up and East is to the left. The images were taken in the $R$ band.}
\label{fc_novae02_fig}
\end{figure*}

\bsp

\label{lastpage}

\end{document}